\documentstyle[12pt,cite]{article}
\jot = 1.5ex\catcode`\@=11\renewcommand{\thefootnote}{\fnsymbol{footnote}}
\begin{document}\begin{titlepage}

\hfill{DFPD/95/TH/42}

\hfill{hep-th/9602174}

\hfill{To appear in Phys. Rev. Lett.}

\hfill{28-Feb-1996}

\vspace{1cm}

\centerline{\large{\bf NONPERTURBATIVE RENORMALIZATION GROUP EQUATION}}

\vspace{0.5cm}

{\centerline{\large{\bf AND BETA FUNCTION IN N=2 SUSY YANG-MILLS}}}

\vspace{1.5cm}

{\centerline{\sc Giulio BONELLI and Marco 
MATONE\footnote[3]{Partly
supported by the European Community Research
Programme {\it Gauge Theories, applied supersymmetry and quantum 
gravity}, contract SC1-CT92-0789}}}

\vspace{0.8cm}

\centerline{\it Department of Physics ``G. Galilei'' - Istituto Nazionale di 
Fisica Nucleare}
\centerline{\it University of Padova}
\centerline{\it Via Marzolo, 8 - 35131 Padova, Italy}

\centerline{bonelli@ipdgr4.pd.infn.it   matone@padova.infn.it}

\vspace{2cm}

\centerline{\bf ABSTRACT}

\vspace{0.6cm}

\noindent
We obtain the exact beta function 
for $N=2$ SUSY $SU(2)$ Yang-Mills theory
and prove the nonperturbative Renormalization Group 
Equation 
$$
\partial_\Lambda{\cal F}(a,\Lambda)=
{\Lambda\over \Lambda_0}\partial_{\Lambda_0}{\cal F}(a_0,\Lambda_0)
e^{-2\int_{\tau_0}^\tau {dx
\beta^{-1}(x)}}.
$$

\end{titlepage}

\newpage

\setcounter{footnote}{0}

\renewcommand{\thefootnote}{\arabic{footnote}}

\noindent
{\bf 1.} 
Montonen-Olive duality \cite{MO} and related versions suggest the 
existence of deep structures underlying relevant QFT's.
As a remarkable example the Seiberg-Witten exact 
results about $N=2$ SUSY Yang-Mills \cite{SW1} (see 
\cite{reviews} for reviews and related aspects),
extensively studied in 
\cite{K,AF,C,FP,KLT,DU,D,BL,HO,A,G,MN,B,1,2,IY,STY,EY,H,W,N,DW,M,I,X,AY,D2,IS},
are strictly related to topics such as uniformization theory,
Whitham dynamics and integrable systems.

In the case of $N=2$ SUSY Yang-Mills with compact gauge group $G$, the terms 
in the low-energy Wilsonian effective action with at most 
two derivatives and four fermions are completely described
by the so-called 
prepotential ${\cal F}$ \cite{S} 
\begin{equation}
S_{eff}={1\over 4\pi}{\rm Im}\,\left(\int d^4xd^2\theta d^2\bar\theta
\Phi^i_D\overline \Phi_i+{1\over 2}\int d^4xd^2\theta
\tau^{ij}W_iW_j\right),
\label{01}\end{equation}
where $W_i$ is a vector multiplet,
 $\Phi^i_D\equiv {\partial {\cal F}/\partial \Phi_i}$ 
is the dual of the chiral superfield $\Phi_i$,
$\tau^{ij}\equiv {\partial^2 {\cal F}/\partial \Phi_i \partial \Phi_j}$
are the effective couplings and $i\in [1,r]$ with $r$ the rank of $G$.

The prepotential ${\cal F}$ plays a central role in the theory.
The most important property of ${\cal F}$ 
is holomorphicity \cite{S}. Furthermore, it has been shown in 
\cite{S} that ${\cal F}$ gets perturbative contributions only up to
one-loop. Higher-order terms in the asymptotic expansion
comes as instanton contribution implicitly determined in \cite{SW1}.

We stress that the exact results obtained by Seiberg-Witten concern
the Wilsonian effective action in the limit considered in (\ref{01}). 
In this context it is useful to recall
that when there are no interacting massless particles
the Wilsonian action and the standard generating functional
of one-particle irreducible Feynman diagrams are identical. In the case
of supersymmetric gauge theories the situation is different. In 
particular
due to IR ambiguities (Konishi anomaly) 
the 1PI effective action might suffer from 
holomorphic anomalies \cite{SF}.

An interesting question concerning the Seiberg-Witten theory is whether
using their nonperturbative results it is possible to reconstruct the 
full quantum field theoretical structure. In this context we note that 
in \cite{1}, where a method
to invert functions was proposed, it has been derived a nonperturbative 
equation which relates in a simple way the prepotential and the vevs of the 
scalar fields. It \cite{STY} 
J. Sonnenschein, S. Theisen and S. Yankielowicz conjectured that the 
above relation should be interpreted in terms of RG ideas.

In this letter we will prove this conjecture. In particular
we will obtain the
nonperturbative Renormalization Group Equation (RGE)
and the exact expression for the beta function of $N=2$ SUSY $SU(2)$
Yang-Mills.

Let us denote by $a_i\equiv \langle \phi^i\rangle$ and
$a^i_D\equiv \langle \phi_D^i\rangle$ 
the vevs of the scalar component of the chiral 
superfield. 
For gauge group $SU(2)$ the moduli space of quantum vacua, parameterized by 
$u\equiv\langle{\rm tr}\, \phi^2\rangle$, is 
$\Sigma_3={\bf C}\backslash\{-\Lambda^2,\Lambda^2\}$, 
the Riemann sphere 
$\widehat {\bf C}={\bf C}\cup \{\infty\}$ with punctures at 
$\pm \Lambda^2$ and 
$\infty$, where $\Lambda$ is the dynamically generated scale.
 It turns out 
that \cite{SW1}
\begin{equation}
a_D(u,\Lambda)=\partial_a{\cal F}=
{\sqrt 2\over \pi}\int_{\Lambda^2}^u {dx \sqrt{x-u}\over 
\sqrt{x^2-\Lambda^4}},
\qquad a(u,\Lambda)={\sqrt 2\over \pi}\int_{-\Lambda^2}^{\Lambda^2}
 {dx \sqrt{x-u}\over 
\sqrt{x^2-\Lambda^4}},
\label{14}\end{equation}
where ${\cal F}$ is the prepotential.
 A crucial step in recognizing the 
full QFT structures underlying the Seiberg-Witten 
theory is the fact that \cite{KLT}
(see also \cite{1})
\begin{equation}
\left[{\partial^2\over \partial u^2}+{1\over 4(u^2-\Lambda^4)}\right]a_D=0=
\left[{\partial^2\over \partial u^2}+{1\over 4(u^2-\Lambda^4)}\right]a,
\label{2}\end{equation}
which is the ``reduction'' of the uniformizing equation 
for $\Sigma_3$, the Riemann sphere with punctures 
at $\pm\Lambda^2$ and 
$\infty$ \cite{C,KLT,1}
\begin{equation}
\left[{\partial^2\over \partial u^2} +{u^2+3\Lambda^4\over 
4(u^2-\Lambda^4)^2}\right]\sqrt{\Lambda^4-u^2}{\partial a_D\over \partial u}=0=
\left[{\partial^2\over \partial u^2}  +{u^2+3\Lambda^4\over 
4(u^2-\Lambda^4)^2}\right]\sqrt{\Lambda^4-u^2}{\partial a\over \partial u}.
\label{11bis}\end{equation}
A related aspect concerns the transformation properties of
${\cal F}$. It turns out that \cite{dWKLL,1}
$$
\gamma \cdot {\cal F}(a)=
\widetilde {\cal F}
(\tilde a)={\cal F}(a)+{a_{11}a_{21}\over 2}a_D^2+
{a_{12}a_{22}\over 2}a^2+a_{12}a_{21}aa_D=
$$
\begin{equation}
{\cal F}(a)+{1\over 4} v^t
\left[G_\gamma^t 
\left(\begin{array}{c}0\\1
\end{array}\begin{array}{cc}1\\0\end{array}\right)
G_\gamma - 
\left(\begin{array}{c}0\\1
\end{array}\begin{array}{cc}1\\0\end{array}\right)
\right]v,
\label{b7tris}\end{equation}
where $v=
\left(\begin{array}{c} a_D\\ a 
\end{array}\right)$ and
 $G_\gamma=\left(\begin{array}{c}a_{11}\\a_{21}\end{array}
\begin{array}{cc}a_{12}\\a_{22}\end{array}\right)\in SL(2,{\bf C})$.
Observe that
$\gamma_2\cdot \left(\gamma_1\cdot {\cal F}(a)\right)=
(\gamma_1\gamma_2)\cdot {\cal F}(a)$ and
\begin{equation}
G_\gamma^t 
\left(\begin{array}{c}0\\1
\end{array}\begin{array}{cc}1\\0\end{array}\right)
G_\gamma - 
\left(\begin{array}{c}0\\1
\end{array}\begin{array}{cc}1\\0\end{array}\right)=2
\left(\begin{array}{c}a_{11}a_{21}\\a_{12}a_{21}
\end{array}\begin{array}{cc}a_{12}a_{21}\\a_{12}a_{22}\end{array}\right).
\label{asada}\end{equation}
We stress that if $G_\gamma\in \Gamma(2)$ then
$\widetilde {\cal F} ={\cal F}$, 
that is $\gamma\cdot {\cal F}(a)={\cal F}(\tilde a)$.
The transformation properties of ${\cal F}$
have been obtained for more general cases
in \cite{dWKLL,STY,deWit}. 
Eq.(\ref{b7tris}) implies that 
$2{\cal F}-a\partial_a{\cal F}$ is invariant under $SL(2,{\bf C})$. In 
particular, it turns out that \cite{1}
\begin{equation}
2{\cal F}-a{\partial{\cal F}\over \partial a} =-8\pi ib_1\langle{\rm 
tr}\, \phi^2\rangle,
\label{81}\end{equation}
where, as stressed in \cite{STY,EY},
 $b_1=1/4\pi^2$ is the one-loop coefficient of the beta function.
Relevant generalizations of the nonperturbative relation (\ref{81}) have 
been obtained by J. Sonnenschein, S. Theisen and S. Yankielowicz \cite{STY}
and by T. Eguchi and S.-K. Yang  \cite{EY}. 

We note that the relation (\ref{81}) turns out to be crucial in 
obtaining  Seiberg-Witten theory from the
tree-level Type II string theory in the limit $\alpha'\to 0$ 
\cite{KKLMV}.

In \cite{STY} it has been 
suggested that Eq.(\ref{81}) should be understood in terms of RG ideas.
In particular, it was suggested to consider the LHS of (\ref{81})
as a measure of the anomalous dimension of ${\cal F}$.
Actually we will see that 
$\langle{\rm tr}\, \phi^2\rangle$ involves the nonperturbative 
beta function in a natural way. This allows us to find the RGE 
for ${\cal F}$.

In order to specify 
the functional dependence of $u$
 we use the notation of \cite{2} by setting $u=\Lambda^2{\cal G}_1(a)$
and $u=\Lambda^2{\cal G}_3(\tau)$
where $\tau=\partial_a^2 {\cal F}$.
Eq.(\ref{2}) implies
\begin{equation}
(1-{\cal G}_1^2)\partial_a^2{\cal G}_1+{a\over 4}
\left(\partial_a{\cal G}_1\right)^3=0,
\label{1}\end{equation}
and by (\ref{81})(\ref{1}) \cite{1,2}
\begin{equation}
\partial_a^3 {\cal F}=
{\left(a \partial_a^2{\cal F}-\partial_a{\cal F}\right)^3
\over 4\left[64\pi^2 b_1^2\Lambda^4+ \left(
{a}\partial_a{\cal F}-
2{\cal F}\right)^2\right]},
\label{gdtf}\end{equation}
which provides recursion relations for the instanton contribution.
By (\ref{14}) we have $a(u=-\Lambda^2,\Lambda)= -i4\Lambda/\pi$
 and  $a(u=\Lambda^2,\Lambda)=4\Lambda/\pi$ so that
the initial conditions for the 
second-order equation (\ref{1}) are
${\cal G}_1(-i4\Lambda/\pi)=-1$ and 
${\cal G}_1(4\Lambda/\pi)=1$.

Eqs.(\ref{81})(\ref{gdtf}) are quite basic for our purpose.
For example, by (\ref{81}) we have \cite{2}
\begin{equation}
{\cal F}(a,\Lambda)={8\pi i b_1}\Lambda^2
a^2\int^a_{4\Lambda/\pi}dx {\cal G}_1(x)x^{-3}
-{i b_1\pi^3 \over 4} a^2,
\label{aaiqwnd}\end{equation}
and \cite{2}
\begin{equation}
\partial_{\hat \tau} \langle {\rm tr}\, \phi^2\rangle=
{1\over 8\pi i b_1}\langle \phi\rangle^2,
\label{icx}\end{equation}
which is the quantum version of the classical relation
$u=a^2/2$. In this context we observe that $\hat\tau=a_D/a$ has
the same monodromy of $\tau$ and  their fundamental domains
differ only for the values of the opening angle at the cusps 
\cite{AFS}. These facts and (\ref{icx}) suggest to
consider $\tau$ and $\hat\tau$ as dual couplings. In particular, 
it should exist a ``dual theory'' with
$\hat\tau$ playing the role of gauge coupling.

As noticed in \cite{STY,EY}, the fact that $\tau=\partial_a^2{\cal 
F}$ is dimensionless implies
\begin{equation}
a(\partial_a{\cal F})_\Lambda+\Lambda(\partial_\Lambda{\cal F})_a=
2{\cal F}.
\label{iuhlk}\end{equation}
Thus, according to (\ref{81}), we have
\begin{equation}
\Lambda\partial_\Lambda{\cal F}=-8\pi ib_1 \langle{\rm tr}\, \phi^2
\rangle.
\label{RG1}\end{equation}

In \cite{2} it has been shown that ${\cal G}_3$
satisfies the equation 
\begin{equation}
2(1-{\cal G}_3^2)^2\left\{{\cal G}_3,\tau\right\}=
-\left(3+{\cal G}_3^2\right)\left(\partial_\tau{\cal G}_3\right)^2,
\label{11tris}\end{equation}
with initial conditions
\begin{equation}
{\cal G}_3(-1)=
{\cal G}_3(1)=-1,\qquad
{\cal G}_3(0)=1.
\label{odiqbis}\end{equation}
The solution of (\ref{11tris}) is
\begin{equation}
u=\Lambda^2{\cal G}_3(\tau)=
\Lambda^2\left\{1- 2\left[\Theta_2(0|\tau)\over 
\Theta_3(0|\tau)\right]^4\right\},
\label{saouihkj}\end{equation}
that by the ``inversion formula''
(\ref{81}) implies \cite{2}
\begin{equation}
2{\cal F}-a{\partial {\cal F}\over \partial a}=
8\pi i b_1\Lambda^2
\left\{2\left[\Theta_2\left(0|\partial_a^2{\cal F}\right)\over 
\Theta_3\left(0|\partial_a^2{\cal F}\right)\right]^4-1\right\},
\label{saouihkjwe}\end{equation}
showing that such a combination of theta-functions acts on 
$\partial^2_a{\cal F}$ as integral operators.

\vspace{1cm}

\noindent
{\bf 2.} Before considering the beta function we observe that 
the scaling properties of $a_D$ and $a$ suggest
to introduce the following notation 
\begin{equation}
\Lambda^{-1}a_D(u,\Lambda)=a_D(v,1)\equiv b_D(v),
\qquad
\Lambda^{-1}a(u,\Lambda)=
 a(v,1)\equiv  b(v),\qquad v\equiv u/\Lambda^2.
\label{scaling}\end{equation}

We now start to evaluate the nonperturbative beta function. 
First of all note that in 
taking the derivative of $\tau$ with respect to $\Lambda$ we have to 
distinguish between $\partial_\Lambda\tau$ evaluated at $u$
or $a$ fixed. We introduce the following notation
\begin{equation}
\beta(\tau)=
\left(\Lambda \partial_\Lambda\tau\right)_u,\qquad 
\beta^{(a)}(\tau)=
\left(\Lambda \partial_\Lambda\tau\right)_a.
\label{betas}\end{equation}

Acting with $\Lambda \partial_\Lambda$ on ${\cal G}_3(\tau)=u/\Lambda^2$, we 
have
\begin{equation}
\beta(\tau){\cal G}_3'(\tau)
=-2{u\over \Lambda^2},
\label{acd1}\end{equation}
so that
\begin{equation}
\beta(\tau)
=-2{{\cal G}_3\over {\cal G}_3'}.
\label{acd2}\end{equation}
Integrating this expression and considering the initial condition
${\cal G}_3(0)=1$ in (\ref{odiqbis}) we obtain
\begin{equation}
\langle {\rm tr}\, \phi^2\rangle_\tau=\Lambda^2e^{-2\int_0^\tau {dx 
\beta^{-1}(x)}},
\label{renormalgroup1}\end{equation}
or equivalently
\begin{equation}
\langle {\rm tr}\, \phi^2\rangle_\tau=\left({\Lambda\over 
\Lambda_0}\right)^2
\langle {\rm tr}\, 
\phi^2\rangle_{\tau_0}
e^{-2\int_{\tau_0}^\tau {dx 
\beta^{-1}(x)}}.
\label{renormalgroup1bis}\end{equation}
Using once again the relation (\ref{81}) we obtain
\begin{equation}
(a\partial_a-2){\cal F}(a,\Lambda)=
8\pi i b_1\Lambda^2e^{-2\int_0^\tau {dx
\beta^{-1}(x)}},
\label{renormalgroup2}\end{equation}
or equivalently
\begin{equation}
(a\partial_a-2){\cal F}(a,\Lambda)=\left({\Lambda\over \Lambda_0}\right)^2
(a_0\partial_{a_0}-2){\cal F}(a_0,\Lambda_0)
e^{-2\int_{\tau_0}^\tau {dx
\beta^{-1}(x)}},
\label{renormalgroup2bis}\end{equation}
which provides the anomalous dimension of ${\cal F}$.
Note that in (\ref{renormalgroup2bis}) we used the notation
$a_0$ to denote $a$ at $\tau_0\equiv\tau(\Lambda_0)$.
By Eqs.(\ref{renormalgroup2})(\ref{renormalgroup2bis})
and (\ref{iuhlk}) we obtain the nonperturbative
Renormalization Group Equation
\begin{equation}
\partial_\Lambda{\cal F}=
-8\pi i b_1\Lambda e^{-2\int_0^\tau {dx
\beta^{-1}(x)}},
\label{renormalgroup3}\end{equation}
that is
\begin{equation}
\partial_\Lambda{\cal F}(a,\Lambda)=
{\Lambda\over \Lambda_0}\partial_{\Lambda_0}{\cal F}(a_0,\Lambda_0)
e^{-2\int_{\tau_0}^\tau {dx
\beta^{-1}(x)}}.
\label{renormalgroup3bis}\end{equation}
We note that, due to the $\tau(\Lambda)$ dependence,
 this equation is highly nonlinear reflecting its nonperturbative 
nature.

\vspace{1cm}

\noindent
{\bf 3.} We now start in deriving from Eq.(\ref{gdtf}) 
an alternative expression for the beta function.
Let us consider the differentials
\begin{equation}
d\tau=(\partial_\Lambda\tau)_a d\Lambda+(\partial_a \tau)_\Lambda da=
(\partial_\Lambda \tau)_ud\Lambda+(\partial_u\tau)_\Lambda du,
\label{dtau}\end{equation}
and
\begin{equation}
da=(\partial_\Lambda a)_ud\Lambda+(\partial_u a)_\Lambda du=bd\Lambda 
+\Lambda b' dv=(b-2vb')d\Lambda+\Lambda^{-1}b'du.
\label{da}\end{equation}
Eqs.(\ref{dtau})(\ref{da}) yield
\begin{equation}
(\partial_\Lambda\tau)_u=
(\partial_\Lambda\tau)_a+
(b-2vb')(\partial_a\tau)_\Lambda.
\label{betau2}\end{equation}
By (\ref{iuhlk}) we have 
$\Lambda(\partial_\Lambda\tau)_a=-a(\partial_a\tau)_\Lambda$, so that
\begin{equation}
\beta(\tau)=-2v b'\Lambda(\partial_a\tau)_\Lambda=
2v{b'\over b}\beta^{(a)}(\tau).
\label{beta3}\end{equation}
Let us introduce ${\cal G}$ and $\sigma$ defined 
by $b_D=\partial_b{\cal G}$ and 
$\sigma=\partial_b^2{\cal G}=b_D'/b'$.
By a suitable rescaling of (\ref{gdtf}), it follows that
$(\partial_b\sigma)_\Lambda=
{1/[2\pi i {b'}^3(1-v^2)]}$.
On the other hand ${\cal G}=\Lambda^{-2} {\cal 
F}$ and $\sigma=\tau$, so that
\begin{equation}
(\partial_b\tau)_\Lambda={1\over 2\pi i {b'}^3(1-v^2)}.
\label{beta4}\end{equation}

Being $\Lambda(\partial_\Lambda\tau)_a=-b(\partial_b\tau)_\Lambda$,
we have
\begin{equation}
\beta(\tau)=
{v\over \pi i {b'}^2(v^2-1)},
\label{beta6}\end{equation}
and
\begin{equation}
\beta^{(a)}(\tau)=
{b\over 2\pi i {b'}^3(v^2-1)}.
\label{beta5}\end{equation}
By (\ref{14})(\ref{saouihkj})(\ref{scaling})(\ref{beta6})(\ref{beta5}) and 
using Riemann's theta 
relation
 $\Theta_3^4=\Theta_2^4+\Theta_4^4$,
where $\Theta_i\equiv \Theta_i(0|\tau)$,
we obtain
\begin{equation}
\beta(\tau)= {2\pi i\left(\Theta_4^4-\Theta_2^4\right)\over
(\Theta_2^8-\Theta_2^4\Theta_3^4)\left[
\large{\int_{-1}^1} dx \sqrt{{1\over 
(x^2-1)(x\Theta_3^4+\Theta_2^4-\Theta_4^4)}}
\right]^2},
\label{beta8}\end{equation}
and
\begin{equation}
\beta^{(a)}(\tau)=
{2\pi i \large{\int_{-1}^1} dx\sqrt{{x\Theta_3^4+\Theta_2^4-\Theta_4^4
\over x^2-1}}\over 
(\Theta_2^8-\Theta_2^4\Theta_3^4)\left[
\large{\int_{-1}^1} dx \sqrt{{1\over 
(x^2-1)(x\Theta_3^4+\Theta_2^4-\Theta_4^4)}}
\right]^3}.
\label{beta7}\end{equation}

Let us discuss some properties of $\beta(\tau)$ and $\beta^{(a)}(\tau)$. 
First of all by (\ref{beta5}) it follows that
$\beta^{(a)}(\tau)$ is nowhere vanishing. This is a consequence
of the fact that $|b|$ has a lower bound that, as noticed
in \cite{FerrariBilal}, is given 
by $b(0)\sim 0.76$. Both $\beta(\tau)$ and $\beta^{(a)}(\tau)$
 diverge at $u=\pm \Lambda^2$ where  dyons and monopoles are massless.
This happens at $\tau\in{\bf Z}$, corresponding to a divergent
 gauge coupling constant.

By (\ref{beta6}) the $\beta(\tau)$ function is vanishing at
$u=0$. We can found the corresponding values of $\tau$ by  
(\ref{saouihkj}). On the other hand, by uniformization theory we know 
that $u=0$ corresponds to $\tau_n=(i+2n+1)/2$, $n\in {\bf Z}$.

As a byproduct of our investigation we observe that (\ref{acd2}) 
and (\ref{beta8}) yield
\begin{equation}
\Theta_2'\Theta_3 -\Theta_2\Theta_3'=
{\Theta_2^5\Theta_3-\Theta_2\Theta_3^5\over 8\pi i}
\left[{\int_{-1}^1} dx \sqrt{{1\over 
(x^2-1)(x\Theta_3^4+\Theta_2^4-\Theta_4^4)}}\right]^2,
\label{thetarelation}\end{equation}
where $\Theta_i'\equiv \partial_\tau\Theta_i(0|\tau)$.

We note that, in a different context, an expression for the beta function was
derived in \cite{NSVZ} whereas
very recently J. Minahan and D. Nemeschansky \cite{MinNem},
using different techniques,
obtained an expression for the beta function which has the same critical points
of $\beta(\tau)$ in (\ref{beta8}).
If one identifies 
(up to normalizations) $\beta(\tau)$ with that in 
\cite{MinNem} one obtains a relation involving the four 
$\Theta_i$'s (including $\Theta_1$).

The beta function has also a geometrical interpretation.
To see this we use 
the Poincar\'e metric on $\Sigma_3$ expressed in terms of vevs in 
\cite{2}. In terms of $\beta$ we have
\begin{equation}
ds_P^2= \left|{\beta\over 2v{\rm Im}\, \tau}\right|^2|du|^2
=e^{\varphi}|du|^2,
\label{BetaLiouville}\end{equation}
so that $\beta/v$ is the chiral block of the Poincar\'e metric.
We observe that
(\ref{2}) is essentially equivalent to the Liouville equation
$2\partial_u\partial_{\bar u}\varphi=e^\varphi$
(see for example \cite{mm3}).

An important aspect of the Seiberg-Witten theory concerns
the structure of the critical curve ${\cal C}$ on which
${\rm Im}\,a_D/a=0$.  The structure and the role of this curve have been
studied in \cite{SW1,F,AFS,LindstromRocek,2,FerrariBilal}.
In particular, in \cite{2},
  using the Koebe 1/4-theorem and Schwarz's lemma, 
inequalities involving the correlators and 
$\Lambda\partial_\Lambda{\cal F}=-8\pi i u$ have been obtained.
Expanding the beta function in the regions of weak and strong
coupling
one has to consider Borel summability for which 
the inequalities in \cite{2} should provide estimations for convergence 
domains.

Finally we observe that the way the results in this paper have been 
obtained suggest
an extension to more general cases.

\vspace{.8cm}

It is a pleasure to thank P.A. Marchetti and M. Tonin for
useful discussions.

\end{document}